\documentclass{WileyMSP-template}
\usepackage{float}
\usepackage{cite}
\usepackage{amsmath, amsfonts, amssymb}
\usepackage{ragged2e}
\begin{document}

\pagestyle{fancy}
\justifying

\title{HiLAB: A Hybrid Inverse-Design Framework}

\maketitle

\author{Reza Marzban(1)}
\author{Hamed Abiri(1)}
\author{Raphaël Pestourie(2)*}
\author{Ali Adibi(1)*}

\dedication{}

\begin{affiliations}(1) School of Electrical and Computer Engineering, Georgia Institute of Technology, Atlanta, GA, USA \\
(2) School of Computational Science and Engineering, Georgia Institute of Technology, Atlanta, GA, USA \\

*Corresponding Authors Email: rpestourie3@gatech.edu, ali.adibi@ece.gatech.edu
\end{affiliations}
\begin{abstract}
\textbf{HiLAB} (\textbf{H}ybrid \textbf{i}nverse-design with \textbf{L}atent-space learning, \textbf{A}djoint-based partial optimizations, and \textbf{B}ayesian optimization), the new paradigm for inverse design of nanophotonic structures presented in this paper, is a framework that combines early-terminated (\emph{partial}) topological optimization (TO), a Vision Transformer–based variational autoencoder (VAE), and a Bayesian search strategy to address multifunctional photonic device design. By halting adjoint-driven TO well before full convergence—and randomizing physical parameters such as thickness and period—we generate a diverse set of freeform photonic configurations at a significantly reduced computational cost. These reasonably good structures are then embedded in a compact latent space by the VAE, enabling the \emph{Bayesian optimization } to \emph{jointly} optimize both the structural pattern (geometry) and physical hyperparameters (thickness, period) of metaphotonic devices. The learned latent space further facilitates  (\emph{by transfer learning}) for new objectives or constraints—such as fabrication tolerances—without necessitating a fresh round of partial TOs. Instead, the trained VAE can be reused, with modifications made only to the acquisition function in the Bayesian optimization to accommodate alternate design goals. Compared to conventional TO pipelines that typically converge to local optima, HiLAB systematically leverages uncertainty-guided sampling to explore close-to-global optima or substantially improved local solutions while reducing electromagnetic simulation requirements. Even accounting for training overhead, HiLAB reduces the total number of full electromagnetic simulations by an order of magnitude, substantially accelerating the exploration of high-performance, fabrication-friendly optical devices. As a proof of concept, we apply HiLAB to the design of an achromatic beam deflector operating at red (660\,nm), green (550\,nm), and blue (470\,nm) wavelengths. The resulting structure achieves balanced diffraction efficiencies—25.2\% for red, 24.7\% for green, and 24\% for blue—while significantly mitigating chromatic aberrations, considerably surpassing all existing demonstrations. Beyond this specific application, HiLAB provides a powerful and generalizable strategy for enforcing multi-parameter constraints, achieving robust designs, and enabling rapid adaptation in designing next-generation nanophotonic and metamaterial devices.
\end{abstract}



\section{Introduction}

The development of new inverse-design techniques, empowered by topological optimization (TO), has recently revolutionized the way we engineer nanophotonic and metamaterial structures by algorithmically shaping device geometries to achieve specific optical responses ~\cite{kuznetsov2024roadmap,li2022empowering,lin2022end,pestourie2020assume,molesky2018inverse}. This paradigm has yielded significant breakthroughs in wavefront shaping \cite{sell2017large,jiang2019global,gershnabel2022reparameterization}, lensing and imaging \cite{li2022inverse,lin2021end}, nonlinear optics \cite{muhammad2025second}, and optical neural networks \cite{fu2024optical,zarei2020integrated,poordashtban2023integrated}. Among the most established inverse-design methods is \emph{gradient-based TO}, which iteratively refines a continuous refractive-index distribution to maximize a desired figure of merit (FoM)\cite{molesky2018inverse,lin2019topology}. However, such methods are prone to convergence toward local optima, requiring hundreds of random restarts from different initial conditions to adequately sample the design landscape~\cite{bendsoe2013topology}. In addition, they typically fix crucial physical parameters—such as thickness or period—based on prior knowledge and/or assumptions about which physical modes or resonances might be supported. While this approach can be effective when parameters are either well understood or experimentally guided, it risks overlooking device geometries whose optimal performance only emerges under alternative or non-intuitive regimes. These drawbacks become especially pronounced in designing \emph{multifunctional} devices (e.g., multi-wavelength or multi-polarization)~\cite{yu2025angle}, demanding simultaneous optimization of performance across multiple spectral or polarization states. A second class of inverse-design approaches, \emph{shape optimization}, more explicitly addresses physical parameters by describing devices in terms of relatively simple geometric primitives (e.g., elliptical or rectangular pillars) whose dimensions and positions can be systematically varied. Such approaches have successfully been applied to metasurface deflectors \cite{jiang2019global}, polarization controllers \cite{gershnabel2022reparameterization}, and mode converters \cite{liao2024low}, among others \cite{dainese2024shape}. In addition, the simpler geometries produced by shape optimization are generally easier to fabricate than freeform layouts obtained from TO approaches. Nonetheless, restricting geometries to basic shapes can overlook complex freeform solutions vital for broadband or multifunctional tasks. Thus, while shape optimization systematically accounts for certain material or dimensional parameters, its limited geometric degrees of freedom often yield suboptimal designs when intricate 2D patterns are essential. A third category of inverse-design methods employs \emph{deep learning}, leveraging neural networks trained on large datasets of device layouts and their corresponding optical responses \cite{kiarashinejad2020deep,liu2018generative,an2019deep}. This data-driven framework has inspired innovative demonstrations, such as inverse-designed color filters \cite{hemmatyar2019full} and spectrally selective metasurfaces \cite{kiarashinejad2020deep,zandehshahvar2023metric}. In principle, once trained, these models can rapidly propose candidate structures without re-running exhaustive simulations. In practice, however, gathering thousands of labeled examples can be computationally expensive, especially since many randomly sampled structures may exhibit poor performance. Furthermore, most deep-learning pipelines rely on relatively simple device parameterizations—often shape-based—because fully freeform representations would require an excessively large number of design parameters, thus demanding even more extensive training data. From a physics perspective, such simplifications can forgo the opportunity to exploit mutual coupling or hybrid modes within complex freeform geometries, which often lead to higher performance. As a result, these methods may overlook richer design spaces that enable coupling intricate geometries with varying physical parameters, a limitation that grows more severe in multifunctional contexts.

Compounding these conceptual hurdles are practical constraints related to electromagnetic (EM) solver speeds and robust device design. For instance, commercial finite-difference time-domain (FDTD) solvers (e.g., Ansys Lumerical) become prohibitively expensive for large-area devices with subwavelength resolution (e.g., millimeter-scale devices at visible wavelengths). Meanwhile, all listed inverse-design techniques, especially purely artificial intelligence (AI)-driven approaches, require large, high-quality training sets; yet random large-scale structures frequently fail to deliver meaningful performance, complicating dataset generation \cite{pestourie2023physics}. Finally, ensuring reliable device operation under inevitable fabrication imperfections remains particularly challenging for broadband or multifunctional designs. 

One prominent example of these challenges is the design of multi-wavelength achromatic beam deflectors, which must steer multiple visible wavelengths at a single angle. Such devices have considerable appeal in display and augmented-reality systems \cite{li2022ultracompact,bonod2016diffraction}, where even moderate overall efficiency can suffice for human perception, but color fidelity remains crucial to avoid noticeable distortions. They are also valuable in hyperspectral imaging or multi-spectral sensing platforms, where correcting chromatic aberrations in a single metasurface can greatly simplify the optical setup. A recent study that applied TO to engineer a multi-wavelength achromatic deflector reported both low and uneven deflection efficiencies (e.g., 22.7\% for red, 14.3\% for green, and 7.1\% for blue)\cite{choi2024multiwavelength}. While some applications may tolerate lower total efficiencies, the \emph{unbalanced} performance across red, green, and blue channels can cause conspicuous color distortion—particularly when only subsets of the wavelengths are used or combined in various proportions—and ultimately undermine the achromatic functionality of the device.

In response to these obstacles, we present in this paper \textbf{HiLAB} (\textbf{H}ybrid \textbf{i}nverse-design with \textbf{L}atent-space learning, \textbf{A}djoint-based partial optimizations, and \textbf{B}ayesian optimization), a framework that merges the versatility of TO with the search efficiency and dimensionality reduction offered by modern AI tools. Rather than performing fully converged TO at a fixed thickness or period of a metasurface, we \emph{truncate} each TO run early and randomly vary key physical parameters, producing a compact dataset of reasonably good (though not optimal) freeform designs. We then train a variational autoencoder (VAE) that leverages a pre-trained Vision Transformer (ViT) for robust feature extraction, enabling it to learn a low-dimensional representation of these partial solutions—including fine-scale freeform details—while drastically reducing the computational cost of subsequent explorations. Finally, a Bayesian optimizer—requiring on the order of 300 evaluations—operates within this learned latent space, enabling \emph{multifunctional} inverse design under practical constraints. As a concrete demonstration, our framework targets a multi-wavelength achromatic beam deflector that preserves identical steering angles for red, green, and blue visible light in a single-layer metasurface. By systematically sampling both freeform geometries and physical parameters, we uncover device configurations and physical modes that remain elusive in conventional pipelines. Inverse design of such an important device with conventional TO typically requires 200 iterations per run and 70 restarts to probe different basins of attraction, resulting in a total simulation budget of 14,000 forward EM solves. If fabrication robustness is considered, conventional robust TO methods typically require additional gradient evaluations for multiple variations of the nominal structure (such as eroded and dilated patterns) \cite{sell2019adjoint,choi2024multiwavelength,sell2017large}, thereby considerably increasing the standard computational workload. By contrast, our method requires only 1050 simulations to train a VAE model and 350 additional evaluations for Bayesian optimization (BO), achieving considerably better performance with just 1400 simulations—more than an order-of-magnitude reduction in cost~\cite{pestourie2025fastapproximatesolversmetamaterials}, agnostic to the choice of underlying solver.
Furthermore, the trained VAE can be easily repurposed for new objectives—such as fabrication tolerances—by simply adjusting the BO acquisition function. Consequently, the need to rerun partial TO is minimized, dramatically reducing the total number of EM simulations. Beyond this specific example, our hybrid approach—combining partial TO, VAE-based dimensionality reduction, and BO—shows significant promise for optimally designing the next generation of robust, large-scale multifunctional nanophotonic devices.
\section{Methodology}
\subsection{Metaphotonic Structure for Optimization}
\label{subsec:metaphotonic_structure_for_optimization}

To demonstrate the effectiveness of our HiLAB inverse-design framework, we consider optimizing a periodic bilayer metasurface composed of co-patterned layers of titania (TiO\textsubscript{2}) and silica (SiO\textsubscript{2}), both deposited on a fused silica substrate, as illustrated in Figure~\ref{fig:concept}(a). The TiO\textsubscript{2} layer serves as the principal design domain and is subdivided into a \(256 \times 128\) binary pixels, a pattern identically replicated in the SiO\textsubscript{2} layer above. To achieve spectrally aligned (transmitted) beam deflection at an identical angle for wavelengths of 660 nm (red), 550 nm (green), and 470nm (blue), we strategically select diffraction orders of 5, 6, and 7, respectively, and subsequently optimize the metasurface structure to maximize their transmission efficiencies \cite{choi2024multiwavelength}. This diffraction-order selection simultaneously satisfies, for the three wavelengths, the classical grating relation:
\begin{equation}
n_t \sin\theta_t = n_i \sin\theta_i + m \frac{\lambda}{\Lambda},
\label{eq:grating}
\end{equation}
where \(n_t\) (\(n_i\)) represents the refractive index of transmitted (incident) medium; \(\theta_t\) 
 (\(\theta_i\)) denotes the transmitted (incident) angle, \(\lambda\) is the operating wavelength, \(\Lambda\) is the unit-cell period, and \(m\) denotes the diffraction order. Fixing the period of the unit-cell at 5\,µm along the x-axis under normal incidence yields a consistent deflection (or diffraction) angle of approximately 41.3\(^\circ\) for all three wavelengths, effectively minimizing chromatic dispersion and enabling achromatic beam deflection. The period along the \(y\)-axis, \(\Lambda_y\), is treated as a variable design parameter and jointly optimized alongside the device thicknesses (t1 and t2) and in-plane geometry as shown in Figure~\ref{fig:concept}(a).

\subsection{Overall Framework}
\label{subsec:framework}

\begin{figure}[H]
  \includegraphics[width=0.95\linewidth]{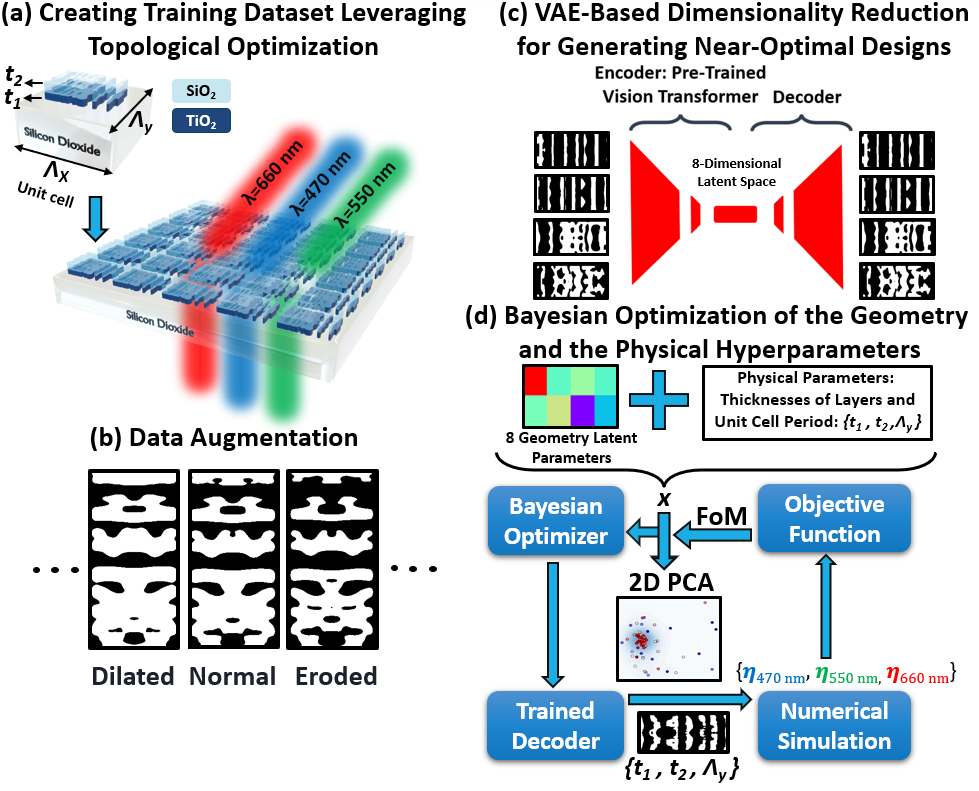}
  \caption{
        Overview of the demonstrated HiLAB pipeline for multi-wavelength metasurface optimization. The metasurface consists of a patterned bilayer comprising SiO\textsubscript{2} and TiO\textsubscript{2}, fabricated on top of a fused silica substrate.
        (\textbf{a}) Multiple partial TO runs under varied physical parameters, including \(t_1\) (TiO\(_2\) thickness), \(t_2\) (SiO\(_2\) thickness), and the unit-cell period \(\Lambda_y\). The period along the \(x\)-direction, \(\Lambda_x\), is fixed at 5\,\(\mu\)m to achieve a 41.3\(^\circ\) deflection angle.
        (\textbf{b}) Morphological data augmentation expands the library without additional EM simulations.
        (\textbf{c}) VAE-based dimensionality reduction (using a pre-trained ViT) learns a compact design manifold.
      (\textbf{d}) BO is performed jointly over the 8-dimensional geometry latent space and physical design parameters—including (\(t_1\), \(t_2\), and \(\Lambda_y\)) to identify metasurface layouts with optimal broadband performance. Each proposed parameter set is decoded into a structural design using the VAE decoder and evaluated through full-wave FDTD simulations. The resulting deflection efficiencies at the three design wavelengths (470 nm, 550 nm, and 660 nm) are used to compute a scalar FoM, defined as the worst-case efficiency across the spectrum: \(\text{FoM} = \min\{\eta_{470}, \eta_{550}, \eta_{660}\}\). This conservative formulation ensures robust spectral performance and serves as the objective function value returned to the Bayesian optimizer. The pipeline incorporates post-decoding smoothing and binarization to ensure that final layouts comply with fabrication constraints such as minimum feature size and pattern co-registration. A 2D Principal Component Analysis (PCA) projection of the 11-dimensional sampled design vectors $x$ is shown to visualize optimizer performance and convergence.
        }
  \label{fig:concept}
\end{figure}

HiLAB is an inverse design framework that couples early-stage partial TO with a deep generative model enabled by a ViT-based VAE. This hybrid scheme balances the physics-based rigor of adjoint-driven optimization with the flexibility to adapt device geometries and physical parameters for a range of objectives. By allowing both geometrical and material properties to vary across truncated TO runs, our method captures a broader spectrum of potential modes and design solutions. The details and different steps of our approach are shown in Figure~\ref{fig:concept} in the context of optimizing a freeform metaphotonic structure. As shown in Figure~\ref{fig:concept}(a), we begin by carrying out multiple \emph{partial} TO runs, each halted after 35 iterations—well before full convergence. The number of iterations has been selected by inspection and trying a few different examples. Although these early truncations do not yield fully optimized devices for our objective function, they select freeform topologies that provide reasonable responses. 
Furthermore, we \emph{randomly sample} key physical parameters (\(\Lambda_y\), \(t_1\), and \(t_2\) in Figure~\ref{fig:concept}(a))  from a normal distribution in each partial run. To observe the fabricability, the mean values for the three parameters are set at $\Lambda_y = 700$~nm, $t_1 = 205$~nm, and $t_2 = 300$~nm, and the standard deviations are 100~nm, 30~nm, and 50~nm, respectively. By populating the design space with distinct structures that provide reasonably good response across different parameter sets, this strategy identifies a more diverse set of solutions than that in conventional workflows, which often fix parameters and simply select the “best” candidate from among many fully converged runs \cite{sell2017large,choi2024multiwavelength,sell2019adjoint}. Critically, we find that only 30 such truncated runs usually suffice to represent a wide range of geometries and materials. In contrast, standard adjoint-based methods often require more than 200 iterations per TO run and 70 such fully converged local optima, imposing a considerably higher computational burden\cite{sell2019adjoint}. In the second step
(Figure~\ref{fig:concept}(b)), we perform data augmentation using morphological transformations \cite{mordvintsev2014opencv} to emulate late-stage TO and enable a richer, fabrication-aware training dataset without requiring additional EM simulations.
\noindent 
In the third step (Figure~\ref{fig:concept}(c)), we train a VAE—whose encoder uses a pretrained ViT—to learn a compact latent representation of the partially optimized structures. The ViT’s global attention mechanism readily accommodates two-dimensional (2D) metasurface patterns in an image-like form, while only the last few layers remain unfrozen to adapt to the metasurface dataset. This VAE projects each freeform topology into a low-dimensional latent space, preserving salient physics-driven features while omitting computationally expensive fine-scale degrees of freedom. By design, the decoder can regenerate high-resolution device layouts from latent vectors, effectively serving as a learned “reparameterization” that includes freeform geometry alongside any randomizable physical parameters (e.g., thickness and period). As a final step (Figure~\ref{fig:concept}(d)), each layout undergoes Gaussian smoothing using a zero-mean Gaussian filter with a fixed standard deviation \(\sigma = 2.0\), followed by binarization with a threshold \(\tau = 0.6\). The Gaussian smoothing is applied via convolution with a 2D isotropic kernel, effectively suppressing subwavelength variations and small-scale noise that are difficult to fabricate. The thresholding operation then converts the smoothed layout into a binary pattern, ensuring compatibility with fabrication constraints and minimum feature size requirements. Once trained, the VAE decoder is embedded in a BO loop, wherein each candidate design is specified by a latent vector plus physical-parameter values as shown in Figure~\ref{fig:concept}(d).\\
 For each proposed point in this joint latent–parameter space, a single FDTD simulation suffices to evaluate the FoM, eliminating the repeated adjoint solves required in conventional TO for every new objective or design condition. This single evaluation markedly lowers the simulation overhead while retaining the ability to explore novel physical regimes (e.g., different thicknesses, multiple wavelengths, or angled incidence). Moreover, if one later wishes to incorporate constraints (e.g., robustness to fabrication imperfections) or change the FoM, only the Bayesian search needs adjusting. This approach thus unifies the key benefits of freeform TO—broad geometric exploration and high efficiency—with a generative model that enables fast re-optimization under evolving constraints, bridging the gap between fully converged adjoint methods and data-driven inverse design techniques.

\subsection{Partial TO and Data Augmentation }
\label{subsec:partial_opt}

\begin{figure}[htbp]
    \centering
    \includegraphics[width=\linewidth]{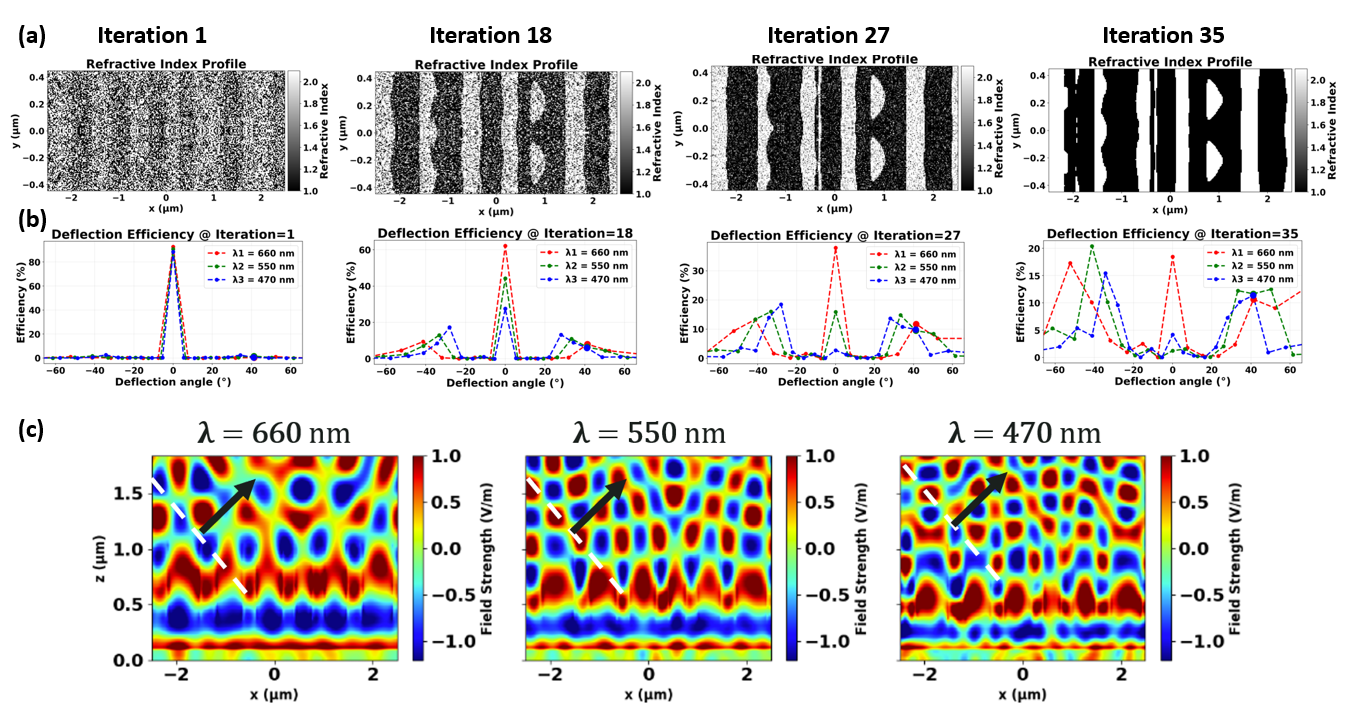}
    \caption{
    Evolution of a metasurface design across partial TO iterations 1, 18, 27, and 35. 
    \textbf{(a)} The refractive-index profile progresses from near-random initialization (iteration~1) to a predominantly binarized pattern (iteration~35). 
    \textbf{(b)} The corresponding deflection efficiencies at red (660\,nm), green (550\,nm), and blue (470\,nm) wavelengths 
    are plotted as functions of outgoing angle. The target (design) deflection angle is \(41.3^\circ\) for all wavelengths, and 
    each efficiency peak shifts progressively closer to this common angle over successive iterations. 
    \textbf{(c)} Electric-field distributions in the \(x\!z\)-plane for the partially optimized structure (iteration 35) are shown 
    for the three wavelengths. The black arrow highlights the deflected wavefront moving toward the desired output angle.
}
    \label{fig:evolution}
\end{figure}

Our \emph{partial} TO approach, involves truncating each adjoint-based optimization run, well before full convergence, at 35 iterations instead of the usual 150--450 \cite{sell2017large,choi2024multiwavelength,sell2019adjoint,su2020nanophotonic}. The number of iterations is a method hyperparameter that balances design quality against simulation time during training-set generation. Our analysis (Supplementary Figure~S1) indicates that at ~30–40 iterations, both the design performance and the core topology stabilize. Continuing optimization beyond this point yields diminishing returns at substantial computational cost. As shown in Figure~\ref{fig:evolution}(a),the principal design domain (The TiO\textsubscript{2} layer) serves as the primary random design region and is discretized into a \(256 \times 128\) pixels along the \(x\)- and \(y\)-directions, respectively, where the refractive index is modulated during adjoint-based TO. As discussed in Section~\ref{subsec:metaphotonic_structure_for_optimization} and illustrated in Figure~\ref{fig:concept}(a), our optimization goal is to maximize transmission efficiencies for selected diffraction orders at target wavelengths. At the end of optimization, the refractive indices of both TiO\textsubscript{2} and SiO\textsubscript{2} layers are binarized to their respective constituent materials resulting in a co-patterned, fabrication-ready bilayer structure. This ensures compatibility with standard lithographic processes and minimizes overlay errors during fabrication.

To promote smooth, manufacturable patterns and to guide convergence, we apply two regularization techniques at every iteration of the TO process: a cone filter and a progressive binarization function~\cite{sell2017large,bendsoe2013topology,christiansen2021inverse}. The cone filter acts as a spatial smoothing operator, convolving the refractive index distribution with a circular kernel to suppress sub-wavelength noise and discontinuities. Its effective radius is updated at each iteration \(i\) according to a logistic schedule:

\begin{equation}
\text{radius}(i) = r_{\text{min}} + \frac{r_{\text{max}} - r_{\text{min}}}{1 + e^{-k \left(\frac{i}{N} - 0.5\right)}}
,\end{equation}
where \(r_{\text{min}} = 1.0\), \(r_{\text{max}} = 2.5\) (in pixel units), \(N\) is the total number of iterations, and \(k = 15\) controls the steepness of the transition. The lower bound (\( r_{\text{min}} \)) corresponds to essentially no filtering in early iterations, allowing the optimizer to freely explore the design space. The upper bound (\( r_{\text{max}}  \)) ensures that the final pattern excludes features smaller than approximately 60\,nm, matching our fabrication constraints given a pixel size of 25\,nm. This iterative smoothing enables the removal of fragmented features in early stages, while refining meaningful geometric patterns in later stages.

In parallel, we enforce a per-iteration binarization using a hyperbolic tangent projection function whose steepness \(\alpha(i)\) evolves logistically with iteration:

\begin{equation}
\alpha(i) = \alpha_{\text{min}} + \frac{\alpha_{\text{max}} - \alpha_{\text{min}}}{1 + e^{-s \left(\frac{i}{N} - 0.5\right)}}
,\end{equation}
with \(\alpha_{\text{min}} = 1.0\), \(\alpha_{\text{max}} = 8.0\), and \(s = 10\). These parameters are selected based on previous experience in working with freeform metasurfaces, without any parameter search or optimization. This gradual sharpening of the projection avoids early convergence to suboptimal local minima by allowing smooth intermediate structures in early iterations while ensuring convergence to a binary, fabrication-compatible pattern as optimization progresses. Such a schedule facilitates a better trade-off between search flexibility and final realizability, helping the design evolve toward a high-performance, fabricable solution by the final iteration. Together, these two operations—applied at each TO step—maintain physical realism, promote robustness, and ensure the final metasurface layout is fabricable. By combining the short-run adjoint optimization with parameter randomization, we assemble a broad and representative library of candidate metasurface designs—each maintaining the coarse freeform features critical to performance. Figure~\ref{fig:evolution}(a) illustrates this progression, showing the evolution of the topology of a typical metasurface design across TO iterations with corresponding responses and field patterns within the metasurface shown in Figure~\ref{fig:evolution}(b) and~\ref{fig:evolution}(c), respectively. 

\noindent 
Finally, as shown in Figure~\ref{fig:concept}(b), we \emph{augment} each device layout by applying a suite of morphological transformations---specifically Gaussian smoothing, erosion, and dilation---to mimic the geometric transitions typically observed during the \emph{intermediate to final stages} of gradient-based TO \cite{mordvintsev2014opencv,christiansen2021inverse}. This is motivated by the observation that in these later TO stages, the structures tend to evolve from noisy or fragmented configurations toward smoother, more regular patterns, as the algorithm refines features to maximize performance while improving fabricability. To emulate this progression, we apply Gaussian filters with varying standard deviations (\(\sigma \in \{1.5, 2.0, 2.5\}\)) where \(\sigma\) is specified in pixel units, and controls the spatial extent of smoothing in the image domain, followed by morphological operations that simulate erosion (material removal) or dilation (feature expansion). These operations are systematically combined to produce variants such as Gaussian + erosion and Gaussian + dilation, in addition to standalone transformations, thereby capturing a broad range of refined topologies characteristic of late-stage optimization. This augmentation pipeline yields a diverse set of plausible designs without requiring any additional partial TO runs that require more extensive EM simulations. Starting with 30 partially optimized designs through TO, we generate different structured combinations of augmentation parameters and then fill out the dataset to 4000 samples using randomly selected transformation types and parameters. The resulting collection reflects realistic trajectories through the design space, encompassing both partially optimized devices and morphology-informed variants of likely high-performance solutions to train our VAE (Section~\ref{subsec:VAE}). By embedding the morphological patterns characteristic of late-stage optimization, this augmentation strategy enhances model generalization and facilitates exploration of high-quality solutions in the downstream inverse design.

\subsection{Vision Transformer VAE for Dimensionality Reduction}
\label{subsec:VAE}
To handle the high-dimensional design patterns obtained from early-terminated TO, we train a VAE \cite{doersch2016tutorial} whose encoder leverages a pretrained ViT \cite{dosovitskiy2020image}. Specifically, we select the publicly available model (\texttt{vit-base-patch16-224-in21k})\cite{wu2020visual,deng2009imagenet} and freeze all but the final two transformer blocks. This partial-thaw strategy~\cite{howard2018universal,zeiler2014visualizing} strikes an optimal balance between retaining the robust, general-purpose features learned from a large dataset (ImageNet-21k~\cite{ridnik2021imagenet}) and adapting the model to our specific metasurface geometries. A detailed ablation study, presented in Supplementary Figure~S2, confirms that this approach achieves the lowest reconstruction error (MSE $\approx 0.0048$) compared to both fully frozen and more extensively fine-tuned models, which were prone to lower fidelity or overfitting, respectively. The encoder outputs a 768-dimensional feature vector, derived from the final layer of the ViT, which captures a global embedding of the input image. This 768-dimensional representation is then projected into a \(d\)-dimensional latent space by adding two parallel fully connected layers. Each of these layers is implemented as a single linear transformation without activation, mapping the ViT feature vector directly to a \(d\)-dimensional output. One layer computes the mean vector \(\boldsymbol{\mu} \in \mathbb{R}^d\), and the other computes the log-variance vector \(\log \boldsymbol{\sigma}^2 \in \mathbb{R}^d\), thereby parameterizing a multivariate Gaussian posterior:
\begin{equation}
\boldsymbol{\mu} = W_\mu\,\mathbf{h}, \quad 
\log \boldsymbol{\sigma}^2 = W_{\log\sigma^2}\,\mathbf{h},
\end{equation}
where \(W_\mu, W_{\log\sigma^2} \in \mathbb{R}^{d \times 768}\) are learnable projection matrices. As shown in Figure~\ref{fig:vae_validation}(b), by selecting a compact latent dimension (\(d = 8\)), we strike a balance between expressiveness and regularization, significantly reducing the dimensionality of the design manifold. This architecture incorporates the standard ``reparameterization trick''\cite{kingma2013auto}, allowing the network to sample the latent vector \(\mathbf{z} \in \mathbb{R}^8\) from the learned distribution during both training and inference:
\begin{equation}
\mathbf{z} = \boldsymbol{\mu} + e^{0.5\,\log\boldsymbol{\sigma}^2} \odot \boldsymbol{\epsilon}, 
\quad \boldsymbol{\epsilon} \sim \mathcal{N}(\mathbf{0}, \mathbf{I}),
\label{eq:reparameterization}
\end{equation}
where \(\odot\) denotes element-wise multiplication and \(\boldsymbol{\epsilon}\) is a standard normal random vector. Equivalently, for each latent dimension \(i \in \{1, \dots, 8\}\), the reparameterization is given by
\begin{equation}
z_i = \mu_i + \sigma_i \cdot \epsilon_i,
\quad \epsilon_i \sim \mathcal{N}(0, 1),
\label{eq:reparam_trick}
\end{equation}
where \(\sigma_i = \exp\left(0.5\, \log \sigma_i^2\right)\) is the standard deviation in the \(i\)-th dimension. Figure~\ref{fig:vae_validation}(b) represents the 8-dimensional latent-space representation of five freeform designs in Figure~\ref{fig:vae_validation}(a). The decoder maps the latent vector \(\mathbf{z} \in \mathbb{R}^8\) back to the image domain (Figure~\ref{fig:vae_validation}(c)) through a multilayer upsampling pipeline. First, a fully connected layer projects \(\mathbf{z}\) into a 16384-dimensional representation, which is then further transformed into a flattened tensor of size \(512 \times 4 \times 8\) through another linear layer. This tensor is reshaped and progressively upsampled via a cascade of five transposed convolutional blocks. Each block uses \(\texttt{kernel\_size} = 4\), \(\texttt{stride} = 2\), and \(\texttt{padding} = 1\), effectively doubling the spatial resolution at each stage. The number of channels decreases sequentially from 512 to 256, 128, 64, 32, and finally 3 in the output layer. Each transposed convolution is followed by batch normalization \cite{ioffe2015batch} and a ReLU \cite{agarap2018deep} activation, except for the final layer, which applies a sigmoid activation function to constrain output pixel intensities to the interval \([0,1]\). This architecture enables the decoder to reliably reconstruct high-contrast metasurface patterns from the compressed (8-dimensional) latent representation, yielding output images at a resolution of \(256 \times 128\) pixels that match the original input size (see reconstructed layouts in Figure~\ref{fig:vae_validation}(c)). Since the input image size of the selected ViT is \(224 \times 224\), we resized the input structure from \(256 \times 128\)  to \(224 \times 224\) using a bilinear interpolation~\cite{kirkland2010bilinear}. Because TO regularization (Section~\ref{subsec:partial_opt}) ensures smooth patterns with a minimum feature size, this resizing preserves the key topological information that governs the optical response.

The VAE is trained by minimizing a composite loss function consisting of three terms. The first is the reconstruction loss, \(L_{\text{recon}}\), which measures the pixel-wise binary cross-entropy between the reconstructed image \(\hat{x}\) and the original input \(x\) (visualized as the reconstruction error in Figure~\ref{fig:vae_validation}(d)):

\begin{figure}
    \centering

  \includegraphics[width=1
  \linewidth]{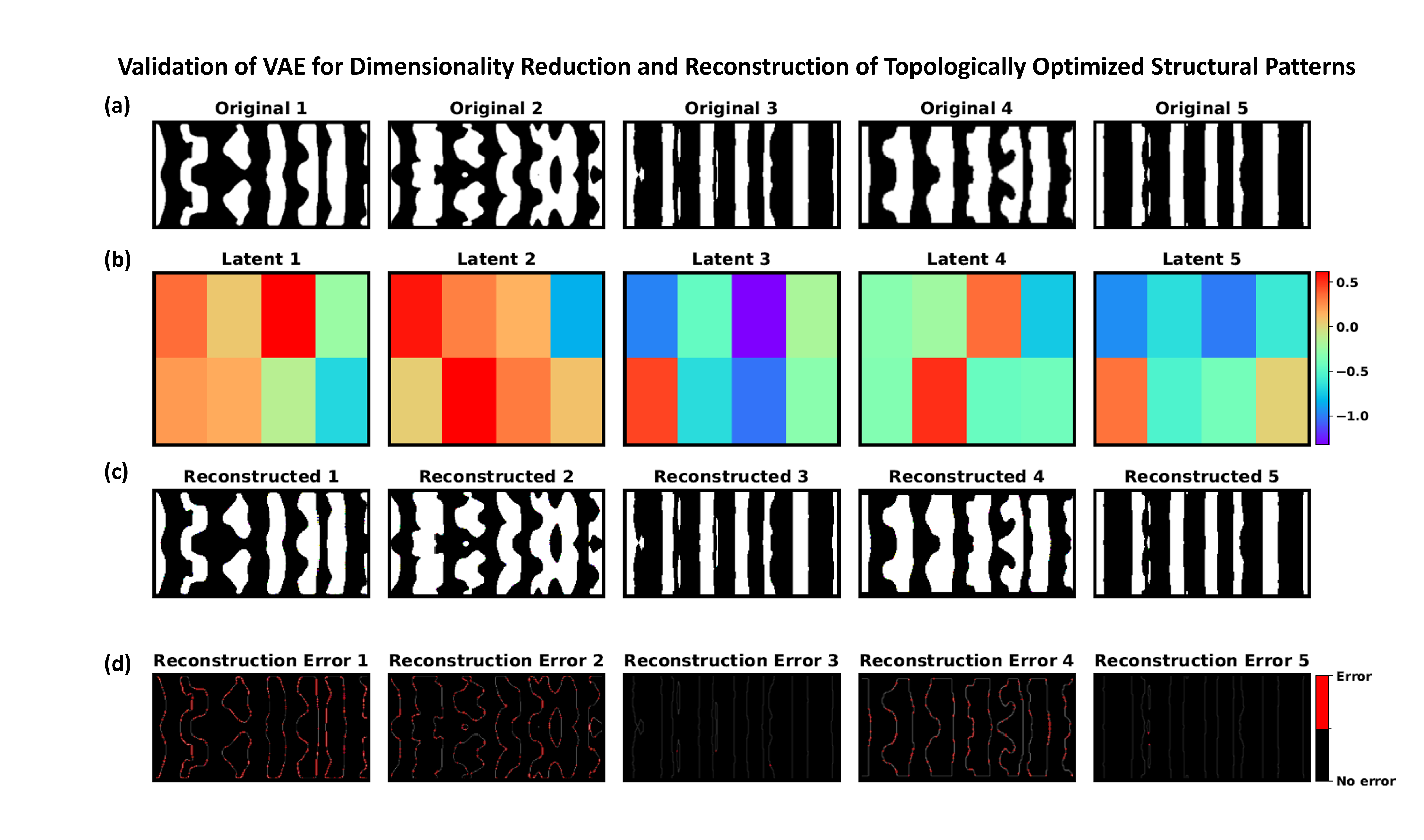}
  \caption{Validation of the ViT-based VAE for dimensionality reduction and reconstruction of partially optimized metasurface designs. Columns correspond to five representative device layouts. Row \textbf{(a)} shows the original topologically optimized structures (Original 1–5) used as input. Row \textbf{(b)} displays the learned latent-space representations (Latent Space 1–5), where each \(8\)-dimensional vector is projected onto a \(2 \times 4\) grid and visualized as a color map. Row \textbf{(c)} presents the VAE-decoded outputs (Reconstructed 1–5), which closely match their respective inputs. Row \textbf{(d)} shows the pixel-wise reconstruction errors (Reconstruction Error 1–5), computed as the absolute difference between the original and reconstructed patterns: \(|\text{Original} - \text{Reconstructed}|\), are near zero across most regions, with noticeable deviations only along the edges of the freeform structures. The minimal error values and close structural correspondence confirm that the VAE effectively captures key freeform features while reducing the input dimensionality by over \textbf{4,000-fold} (\(256 \times 128\) to 8).
  }
    \label{fig:vae_validation}
\end{figure}
\begin{equation}
\label{eq:Lrecon}
L_{\text{recon}} 
= -\sum_{i}\Bigl[x_i\log(\hat{x}_i)
    +\bigl(1 - x_i\bigr)\log\bigl(1 - \hat{x}_i\bigr)\Bigr].
\end{equation}
Second, the Kullback--Leibler (KL) divergence, $L_{\text{KLD}}$, serves as a regularization term encouraging the approximate posterior distribution over $\mathbf{z}$ to align closely with a standard Gaussian prior, thus facilitating smooth and interpretable interpolations within the latent space:

\begin{equation}
\label{eq:LKLD}
L_{\text{KLD}} 
= -\tfrac{1}{2}\sum_{j}\Bigl[\,1 + \log\bigl(\sigma_j^2\bigr) 
    - \mu_j^2 - \sigma_j^2\Bigr],
\end{equation}
where $\mu_j$ and $\sigma_j$ are the mean and standard deviation, respectively, of the latent distribution in dimension $(1\leq j\leq8)$. Third, the binarization penalty, $L_{\text{binarization}}$, promotes outputs toward near-binary states to produce high-contrast metasurface patterns:

\begin{equation}
\label{eq:Lbinarization}
L_{\text{binarization}} = \mathbb{E}\bigl[\hat{x}\,\bigl(1 - \hat{x}\bigr)\bigr].
\end{equation}
Since this term achieves minimal values when $\hat{x}$ approaches binary states (0 or 1), it effectively discourages intermediate grayscale intensities, thereby helping produce patterns that are more readily fabricable. Finally, the total training loss $L_{\text{total}}$ combines these terms:

\begin{equation}
\label{eq:Ltotal}
L_{\text{total}} 
= L_{\text{recon}} 
  + \alpha \,L_{\text{KLD}} 
  + \beta \,L_{\text{binarization}},
\end{equation}
where \(\alpha = 0.02\) and \(\beta = 0.1\) balance the influence of KL divergence and binarization penalties, respectively. These values were selected based on a coarse grid search aimed at minimizing validation reconstruction error while ensuring the latent space remained informative (via KL regularization) and the output patterns remained nearly binary (via binarization loss). Further research could focus on more detailed optimization of these parameters as well. Minimizing \(L_{\text{total}}\) yields latent representations that not only accurately reconstruct the input image but also promote near-binary outputs suitable for practical metasurface fabrication. As shown in Figures ~\ref{fig:vae_validation}(c) and \ref{fig:vae_validation}(d), the reconstructed layouts closely match their respective inputs with minimal error across all samples. These results confirm the ViT-based VAE effectiveness in preserving critical freeform features while achieving over \textbf{4,000-fold} dimensionality reduction.

\begin{figure}[ht]
    \centering
    \includegraphics[width=1\linewidth]{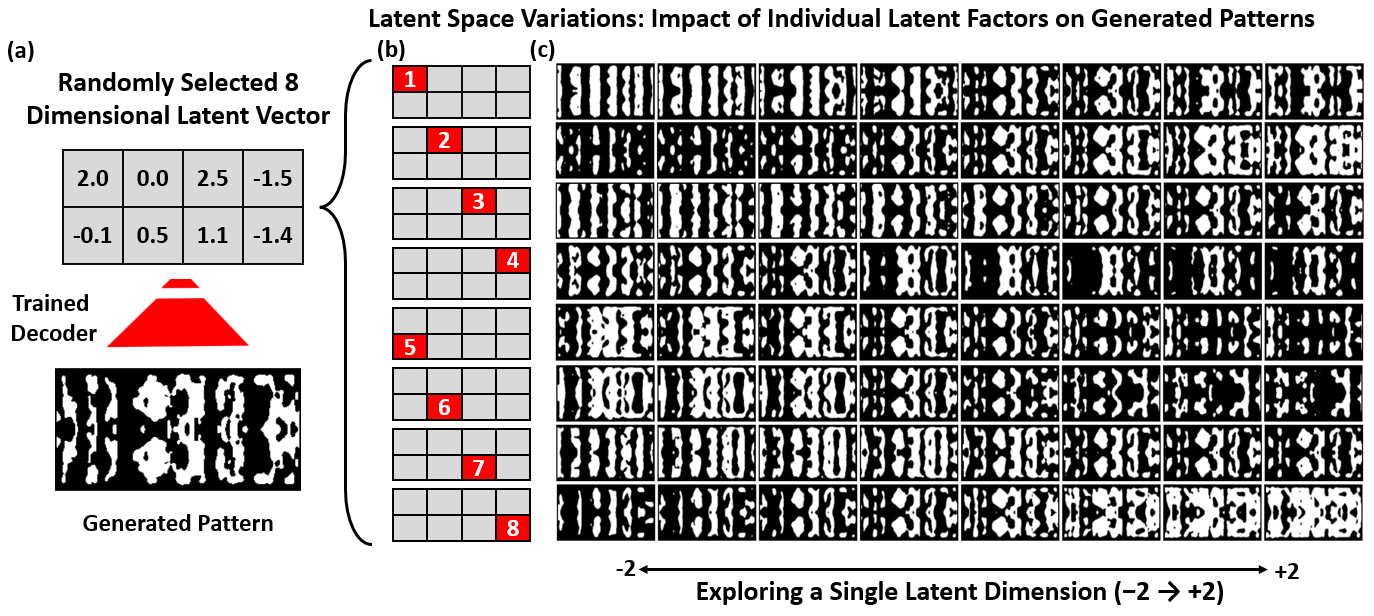}
    \caption{Visualization of how individual elements of the latent vectors influence the generated metasurface patterns. \textbf{(a)} A randomly selected 8-dimensional latent vector \([2.0,\ 0.0,\ 2.5,\ -1.5,\ -0.1,\ 0.5,\ 1.1,\ -1.4]\) is decoded using the trained model to produce a binary pattern. \textbf{(b)} Schematic highlighting which latent component (from 1 to 8, red square) is being varied in each row.
    \textbf{(c)} Binarized outputs generated by sweeping the selected latent component from \(-2.0\) to \(+2.0\), while all other seven component remain fixed. Thus, each row corresponds to a unique latent component and each column to a specific value along that component. The distinct structural variations confirm that the learned latent space is interpretable and disentangled, with different elements controlling different pattern characteristics.}
    \label{fig:vae_var}\end{figure}

As illustrated in Figure~\ref{fig:vae_var}, sweeping each latent vector individually reveals how specific dimensions of the 8-dimensional latent space modulate the generated metasurface patterns. Beginning with a randomly selected latent vector (Figure~\ref{fig:vae_var}(a)), each component is varied across a fixed range while holding the remaining components constant. The active dimension in each case is highlighted in Figure~\ref{fig:vae_var}(b), and the resulting decoded structures are shown in Figure~\ref{fig:vae_var}(c). The systematic and visually distinct changes across rows demonstrate that the VAE has learned a disentangled and interpretable latent representation, with individual variables (latent vector dimensions) governing specific geometric transformations. This property is crucial during the BO phase (Section~\ref{subsec:Bayes}), where we leverage the latent space to refine device performance.

\subsection{VAE-Driven Design with Bayesian Optimization}
\label{subsec:Bayes}

A widely adopted strategy in the inverse design of photonic devices involves constructing a large dataset of \((\text{structure},\,\text{response})\) pairs and training neural networks to map the design space to the response  space~\cite{kiarashinejad2020deep,liu2018generative,an2019deep,hemmatyar2019full,zandehshahvar2023metric}. Although once trained, these models can rapidly predict device performance or suggest candidate designs, their coverage of the design landscape is constrained by the limited size of the training set. Generating such datasets is extremely time-consuming because it requires numerous full-wave EM simulations, and the resulting collection of samples still spans only a small fraction of the overall design space, severely constraining the ability of any data-driven model to generalize effectively.
Here, we address this limitation by combining the VAE approach with BO while appending three real-valued physical parameters: ${t_1}$ and ${t_2}$, representing the thicknesses of the TiO$_2$ and SiO$_2$ layers, respectively, and ${\Lambda_y}$, denoting the unit-cell period along the $y$-axis. These parameters, illustrated schematically in Figure~\ref{fig:concept}(a), not only represent essential fabrication constraints but also serve as critical physical hyperparameters that can enable better designs. Together with the latent vector, these parameters form an 11-dimensional hybrid representation, as shown in Figure~\ref{fig:concept}(d):
\begin{equation}
\mathbf{x} 
\,=\, 
\bigl(\mathbf{z},\,{\Lambda_y},\,{\mathrm{t_1}},\,{\mathrm{t_2}}\bigr),
\label{eq:design_vector}
\end{equation}
Each design \(\mathbf{x}\) is evaluated using a full-wave FDTD simulation, which returns the diffraction efficiencies \(\eta_{470},\, \eta_{550},\, \eta_{660}\) at the target wavelengths. The FoM is then defined as the worst-case efficiency across the three wavelengths:

\begin{equation}
\mathrm{FoM}(\mathbf{x}) 
= \min\{\eta_{470},\, \eta_{550},\, \eta_{660}\} 
= \mathcal{F}\bigl[\mathrm{Decoder}(\mathbf{z});\,{\Lambda_y},\,{\mathrm{t_1}},\,{\mathrm{t_2}}\bigr],
\label{eq:FoM}
\end{equation}
where \(\mathcal{F}[\cdot]\) denotes the full FDTD simulation and post-processing pipeline. Note that the FDTD simulation is performed over the freeform structure that is created by decoding the 8-parameter latent vector and combining it with the 3 physical parameters. This scalar FoM serves as the objective function for BO, guiding the search toward high-efficiency, broadband metasurface designs.
To efficiently navigate the 11-dimensional design space $\mathbf{x}$ defined in Eq.~\eqref{eq:design_vector}, we employ BO with a Gaussian Process (GP) surrogate~\cite{gramacy2020surrogates,mcintire2016sparse}, a method that has proven highly effective for multi-objective optimization of complex metamaterials~\cite{xi2023ultrahigh,chen2023multi}. Unlike methods that exclusively rely on pre-generated, static datasets, our framework sequentially updates the GP surrogate by incorporating new data obtained from FDTD evaluations. At each iteration, a candidate design vector $\mathbf{x}_t$ is decoded into a specific metasurface layout, evaluated via full-wave EM simulation to compute its FoM defined by Eq.~\eqref{eq:FoM}, and then integrated as an observation into the GP surrogate. This iterative closed-loop optimization procedure, illustrated in Figure~\ref{fig:concept}(d), enables systematic exploration and targeted exploitation of the high-dimensional design landscape, dramatically reducing the number of required computationally expensive EM simulations. Formally, the FoM is modeled as a random function drawn from a GP prior with zero mean:
\begin{equation}
    \mathrm{FoM}(\mathbf{x}) \sim \mathcal{GP}\left(0,\,k(\mathbf{x},\mathbf{x}')\right),
    \label{eq:gp_prior}
\end{equation}
where the kernel $k(\mathbf{x},\mathbf{x}')$ explicitly encodes our assumptions about correlations within the combined latent and physical design space. The pairwise distance between design vectors $\mathbf{x}_i$ and $\mathbf{x}_j$ is computed using the Euclidean norm:
\begin{equation}
\mathbf{r}=\|\mathbf{x}_i - \mathbf{x}_j\| = \sqrt{\sum_{d}(x_{i,d} - x_{j,d})^2},
\end{equation}
where the index $d$ runs over all dimensions of the design vector. We then adopt a hybrid covariance kernel composed of two complementary terms: a Matérn 5/2 kernel for smooth but adaptive variation, and a periodic component based on a radial basis function (RBF) modulated by an exponential sine-squared factor\cite{seeger2004gaussian}:
\begin{equation}
    k(\mathbf{x}_i, \mathbf{x}_j) = 
    k_{\mathrm{Mat\acute{e}rn}_{5/2}}(\mathbf{r}) + 
    k_{\mathrm{RBF}}(\mathbf{r}) \exp\left(-\frac{2}{l^2}{\sin^2(\frac{\pi \mathbf{r}}{p})}\right),
    \label{eq:hybrid_kernel}
\end{equation}
where $l$ is the length-scale parameter governing how rapidly correlations between points in the design space diminish with increasing distance, and $p$ is the periodicity parameter controlling the repetition frequency of the kernel's periodic component. Additionally, the Matérn 5/2 kernel component explicitly contributes a flexible smoothness characteristic, defined as:
\begin{equation}
k_{\text{Matérn}_{5/2}}(\mathbf{r}) = 
\sigma^2 \left(1 + \frac{\sqrt{5}\mathbf{r}}{l} + \frac{5\mathbf{r}^2}{3l^2}\right)\exp\left(-\frac{\sqrt{5}\mathbf{r}}{l}\right),
\end{equation}
where $\sigma^2$ controls the variance or amplitude of the function modeled by the GP. This formulation provides an optimal balance between smoothness and local flexibility, allowing the GP surrogate to capture subtle, gradually varying behaviors in addition to localized or potentially repeating structures. 
 
Consequently, this hybrid kernel structure maintains high predictive accuracy even when periodic or complex relationships within the design space are subtle or uncertain. At each optimization iteration, the GP surrogate leverages previously simulated designs to generate predictions of mean performance $\mu_t(\mathbf{x})$ and associated uncertainty $\sigma_t(\mathbf{x})$. We employ an Upper Confidence Bound (UCB) acquisition function to select subsequent designs, balancing the exploration of uncertain areas (high $\sigma_t(\mathbf{x})$) against the exploitation of regions expected to yield high performance (high $\mu_t(\mathbf{x})$):
\begin{equation}
    \mathbf{x}_{t+1} = \arg\max_{\mathbf{x}}\left[\mu_t(\mathbf{x}) + \gamma\,\sigma_t(\mathbf{x})\right],
    \label{eq:acquisition}
\end{equation}
where $\gamma$ explicitly controls the exploration-exploitation balance, with larger values encouraging exploration of less certain regions, and smaller values favoring exploitation of high-confidence, high-performance regions. The chosen design is decoded into a physical metasurface geometry and rigorously evaluated using a full-wave EM simulation, updating the GP surrogate in the process.

\begin{figure}[H]
  \centering
  \includegraphics[width=1\linewidth]{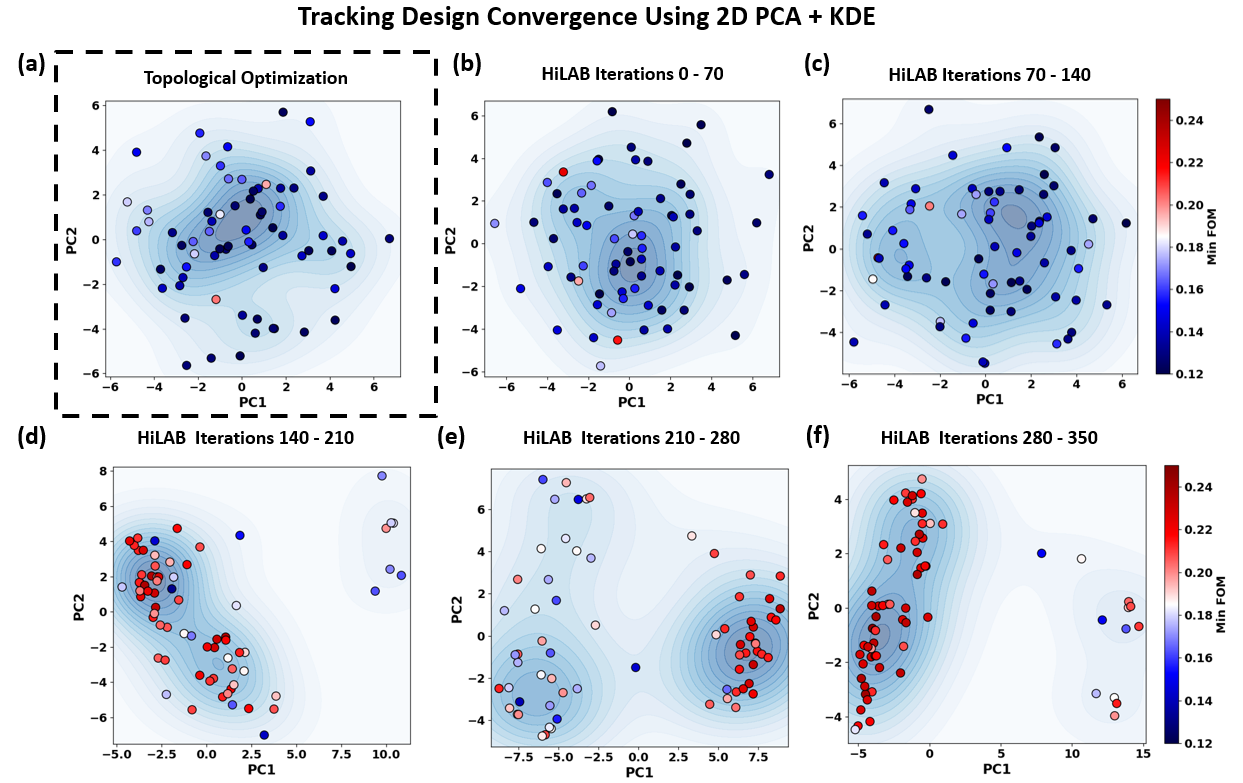}
 \caption{
    2D PCA + Kernel Density Estimation (KDE) visualization of the 11-dimensional design space \(\mathbf{x}\), defined in 
    Eq.~\eqref{eq:design_vector}(composed of 8 latent variables and 3 physical parameters (\(t_1\), \(t_2\), and \(\Lambda_y\))). 
    \textbf{(a)}~TO results from 70 independent runs, each with 
    200 iterations. For each run, the physical parameters were randomly sampled to generate 
    diverse initial conditions, and the final structure was encoded using the trained VAE 
    to extract the 8 latent features. These were then combined with the physical parameters 
    to construct the full design vector \(\mathbf{x}\) for PCA--KDE visualization. The best FoM was 0.188 (minimum efficiency of 18.8\% between three wavelengths).  
    \textbf{(b)--(f)}~Snapshots of our \emph{HiLAB} framework at 70-iteration intervals 
    (0--70, 70--140, 140--210, 210--280, 280--350). Despite requiring only 1{,}400 simulations 
    in total, \emph{HiLAB} discovers 
    compact, high-performing clusters with a peak FoM of 0.247(minimum efficiency of 24.7\% between three wavelengths). 
    KDE contours in each panel reflect the density of sampled designs in the projected PCA space: 
    while \emph{HiLAB} progressively concentrates sampling in promising regions, 
    the broader, more diffuse KDE distribution in panel (a) reflects the randomness and lack 
    of convergence in the baseline TO with randomized physical conditions.
    }
  \label{fig:TOPOLOGICAL_VS_Hilab}
\end{figure}

To visualize \emph{where} the optimizer is sampling inside the
11-dimensional design space~$\mathbf{x}$, we project every evaluated
design onto a two-dimensional principal-component plane and overlay
kernel-density contours colored by their corresponding
$\mathrm{FoM}(\mathbf{x})$.
For better visualization of the design process, we map the designs into a 2D space by training a 2D principal component analysis (PCA) algorithm using FOM as the metric. We also use Kernel Density Estimation (KDE) contours to visualize the density of sampled designs in the projected PCA space. We also implement pure TO-based design for comparison purposes. The details of our design approach using both HiLAB and TO are shown in Figure~\ref{fig:TOPOLOGICAL_VS_Hilab}. Figure~\ref{fig:TOPOLOGICAL_VS_Hilab}(a) presents the results of 70 independent conventional TO runs, each consisting of 200 iterations with randomly selected physical parameters ${{t_1}, {t_2}, {\Lambda_y}}$ for the structure in Figure ~\ref{fig:concept}(a). Each point in the 2D PCA space corresponds to one final TO design. Despite conducting 14,000 EM simulations, conventional TO achieves a modest best-case FoM of 0.188 (among all 70 runs).  The scattered distribution of the design points in the PCA space emphasizes conventional TO’s susceptibility to local optima and sensitivity to initial conditions. In contrast, our HiLAB framework delivered significantly superior performance (FOM = 0.247) using only about 1,400 total simulations—1,050 for initial VAE training and 350 for the GP-driven BO step—representing an order of magnitude reduction in computational demand compared to conventional TO. 
\noindent
 Figures~\ref{fig:TOPOLOGICAL_VS_Hilab}(b)-~\ref{fig:TOPOLOGICAL_VS_Hilab}(f) illustrate how HiLAB progressively refines sampling, converging to compact, high-performing clusters and ultimately achieving a peak FoM of 0.247 (24.7\% minimum deflection efficiency among the three wavelengths). The evolution of the approach toward designs with larger FOM (identified by red dots) at higher iterations is clear from Figures~\ref{fig:TOPOLOGICAL_VS_Hilab} (b)–(f).  Considering fabrication robustness further emphasizes HiLAB's advantages. Conventional robust TO methods typically require additional gradient evaluations at each optimization step for three distinct structural conditions—eroded, nominal, and dilated—to explicitly represent possible fabrication imperfections. These extra evaluations effectively triple the standard computational load~\cite{sell2017large,sell2019adjoint,choi2024multiwavelength}. Such extensive simulation demands arise because even small fabrication inaccuracies, such as edge deviations on the order of $\pm 10\,\mathrm{nm}$, can drastically reduce device efficiency by as much as  50\%~\cite{sell2019adjoint}. Consequently, explicitly incorporating these fabrication-induced variations into the optimization loop guides conventional TO towards more robust and fabrication-tolerant designs. In contrast, HiLAB inherently identifies clusters of high-performing solutions within the latent-physical parameter space. By selectively sampling from these clusters, designers can efficiently pinpoint solutions that are robust against such fabrication imperfections without incurring additional computational overhead from extra gradient evaluations. Furthermore, since HiLAB identifies regions within the design space ($\mathbf{x}$) containing numerous high-performance designs (e.g., as seen in Figure~\ref{fig:TOPOLOGICAL_VS_Hilab}(f)), designers have the flexibility to select from these solutions, choosing designs with features (e.g., patterns with higher or lower aspect ratios) that simplify fabrication without compromising performance
To rigorously validate reproducibility, we conducted the complete HiLAB pipeline five independent times, each involving fresh partial TO dataset generation, VAE retraining, and reinitialization of the BO step. A detailed visualization of one such independent run, which confirms the repeatable discovery of a high-performance design cluster, is provided in Supplementary Figure~S5. All trials consistently exceeded a minimum efficiency of 23.9\%, averaging approximately 24.2\% with a minimal standard deviation ($\sim$0.3\%), clearly confirming HiLAB’s robustness and reproducibility compared to conventional methods.
\begin{figure}[H]
  \centering
  \includegraphics[width=.86\linewidth]{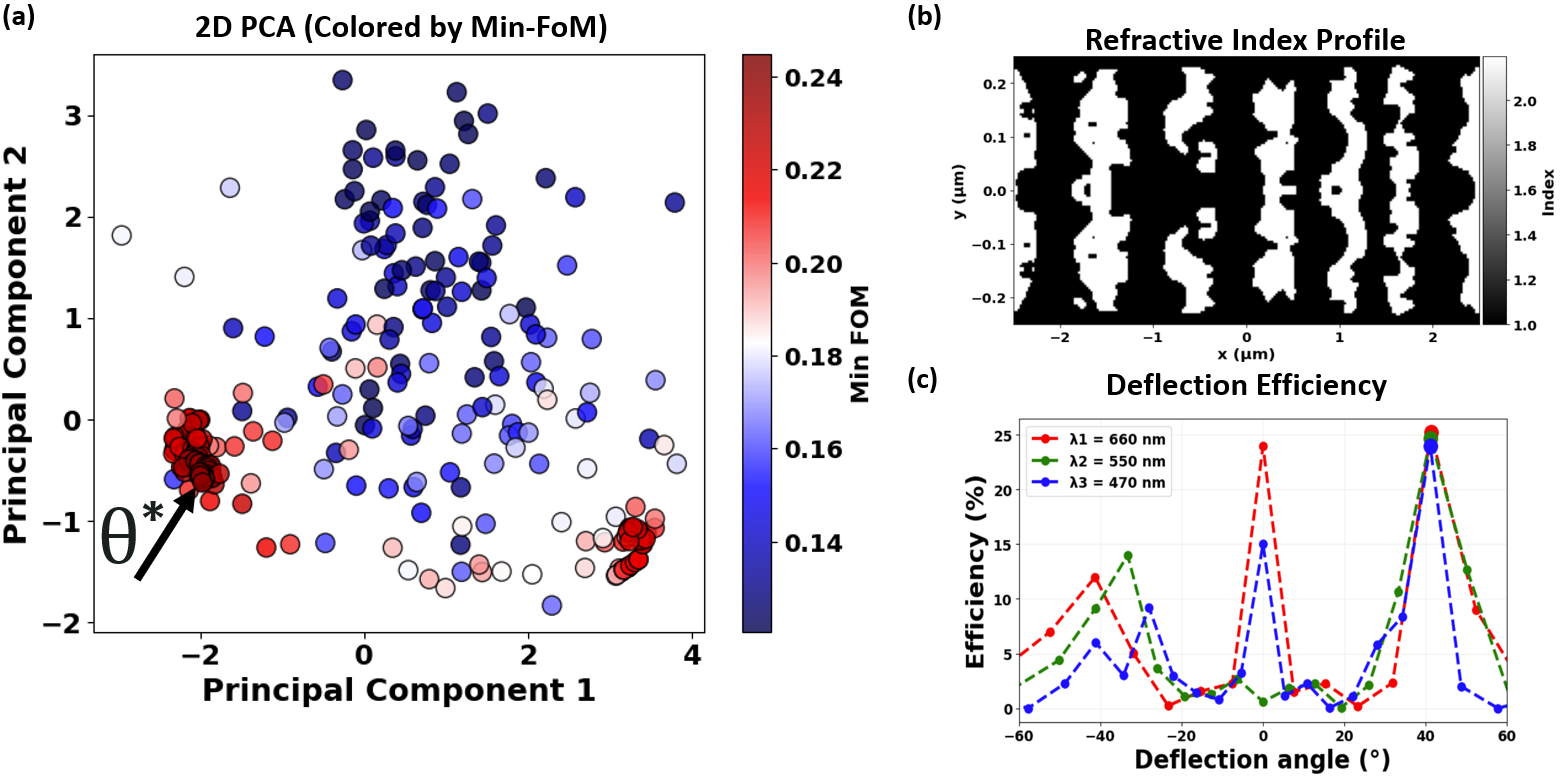}
\caption{
    \textbf{(a)} Projection of all evaluated 11-dimensional design vectors $\mathbf{x}$ onto a two-dimensional PCA space. Each point represents a specific combination of latent geometry parameters and physical hyperparameters, color-coded according to the minimum simulated FoM across wavelengths (660\,nm, 550\,nm, and 470\,nm) up to that iteration. The clustering of the FoM data clearly indicates the systematic exploration of the optimizer towards high-performance regions. A representative optimal design (\(\theta^*\)) is highlighted by the black arrow, with physical parameters: period $\Lambda_y = 500\,\mathrm{nm}$, TiO$_2$ thickness = $196\,\mathrm{nm}$, and SiO$_2$ thickness = $400\,\mathrm{nm}$.
    \textbf{(b)} Refractive-index profile of the optimized metasurface unit cell corresponding to design \(\theta^*\).
    \textbf{(c)} Simulated deflection efficiencies for red (660\,nm), green (550\,nm), and blue (470\,nm) wavelengths, demonstrating the strong multi-wavelength achromatic deflection capability of the device at the target \(41.3^\circ\) angle with an unprecedented efficiency and almost equal efficiency among the three colors.
}   
  \label{fig:PCA}
\end{figure}
\noindent
Figure~\ref{fig:PCA}(a) illustrates a representative result from these trials, highlighting HiLAB’s systematic optimization strategy. A representative optimal design, labeled $\theta^{\ast}$, is clearly positioned within a distinct cluster, achieving remarkable deflection efficiencies of 25.2\%, 24.7\%, and 24.0\% at wavelengths of 660 nm, 550 nm, and 470 nm, respectively. Figure~\ref{fig:PCA}(b) shows the refractive-index profile of the optimized metasurface unit cell corresponding to design $\theta^{\ast}$, and Figure~\ref{fig:PCA}(c) demonstrates its superior multi-wavelength achromatic deflection performance.
By integrating generative dimensionality reduction via VAE encoding—achieving dimensionality reduction greater than 4000-fold—with Bayesian surrogate optimization, the HiLAB framework effectively addresses computational challenges inherent in inverse nanophotonic design. This approach significantly reduces computational cost while offering a highly scalable and broadly applicable solution, extendable to various scientific and engineering inverse-design problems.

To avoid ambiguity, we emphasize that the VAE stochastic behavior during training is entirely separate from, and should not be confused with, its deterministic role during the optimization stage. The generative nature of the training process, driven by stochastic sampling, allows the VAE to learn a unique and continuous latent representation corresponding to the training dataset and that is structured by a Gaussian distribution. Once this specific representation is learned and fixed, we use it as a deterministic decoder from the latent vector to a device geometry. Our method then shifts to a Bayesian optimizer that finds the optimal latent vector $z^*$ (that was not seen during training). The key hyperparameters governing the TO regularization, VAE loss function, and post-processing steps were selected based on established literature and empirical tuning. The rationale for each parameter choice is detailed in Supplementary Section~S3.

\section{Experimental Results and Discussion}

\subsection{Fabrication Process}
One of the optimal multi-wavelength metasurface designs with feature sizes larger than 50 nm (Figure~\ref{fig:fab}(a)) is fabricated on a fused silica substrate that is first thoroughly cleaned and prepared. A thin film of \(\text{TiO}_2\), nominally 187\,nm in thickness, is deposited via thermal evaporation on the substrate. Subsequently, a 300\,nm layer of \(\text{SiO}_2\) is added to serve as the base structure. Hydrogen silsesquioxane (HSQ), a negative-tone electron-beam (e-beam) resist with material properties closely resembling \(\text{SiO}_2\), is employed for patterning, masking, and serving as the top layer of the device using e-beam lithography (EBL). After \(\text{TiO}_2\) deposition, the sample is spin-coated with HSQ and overlaid with a conductive, water-soluble Espacer film to mitigate charging during the e-beam writing step. The HSQ thickness is adjusted to approximately 300\,nm above the \(\text{TiO}_2\) layer. EBL is then used to define the metasurface features, typically in \(200\,\mu\text{m} \times 200\,\mu\text{m}\) blocks. After exposure, the Espacer is removed with deionized (DI) water, and the HSQ resist is developed in 25\% tetramethylammonium hydroxide (TMAH), followed by a DI water rinse. Pattern transfer into the underlying TiO\textsubscript{2} is accomplished through inductively coupled plasma reactive-ion etching (ICP-RIE) using Ar/O\textsubscript{2}/CF\textsubscript{4} chemistry (Ar: argon, O\textsubscript{2}: oxygen, CF\textsubscript{4}: carbon tetrafluoride). Under optimized conditions, the etch proceeds at approximately 150~nm/min. The residual HSQ effectively acts as a hard mask and is subsequently baked to fully convert it into a material with properties resembling SiO\textsubscript{2}. This final step completes the formation of the top-layer nanostructures, yielding the fully realized metasurface. The scanning electron micrograph (SEM) of the fabricated metasurface is shown in Figure~\ref{fig:fab}(b).

\subsection{Multispectral Characterization System and Results}
\begin{figure}[H]
  \centering
  \includegraphics[width=1\linewidth]{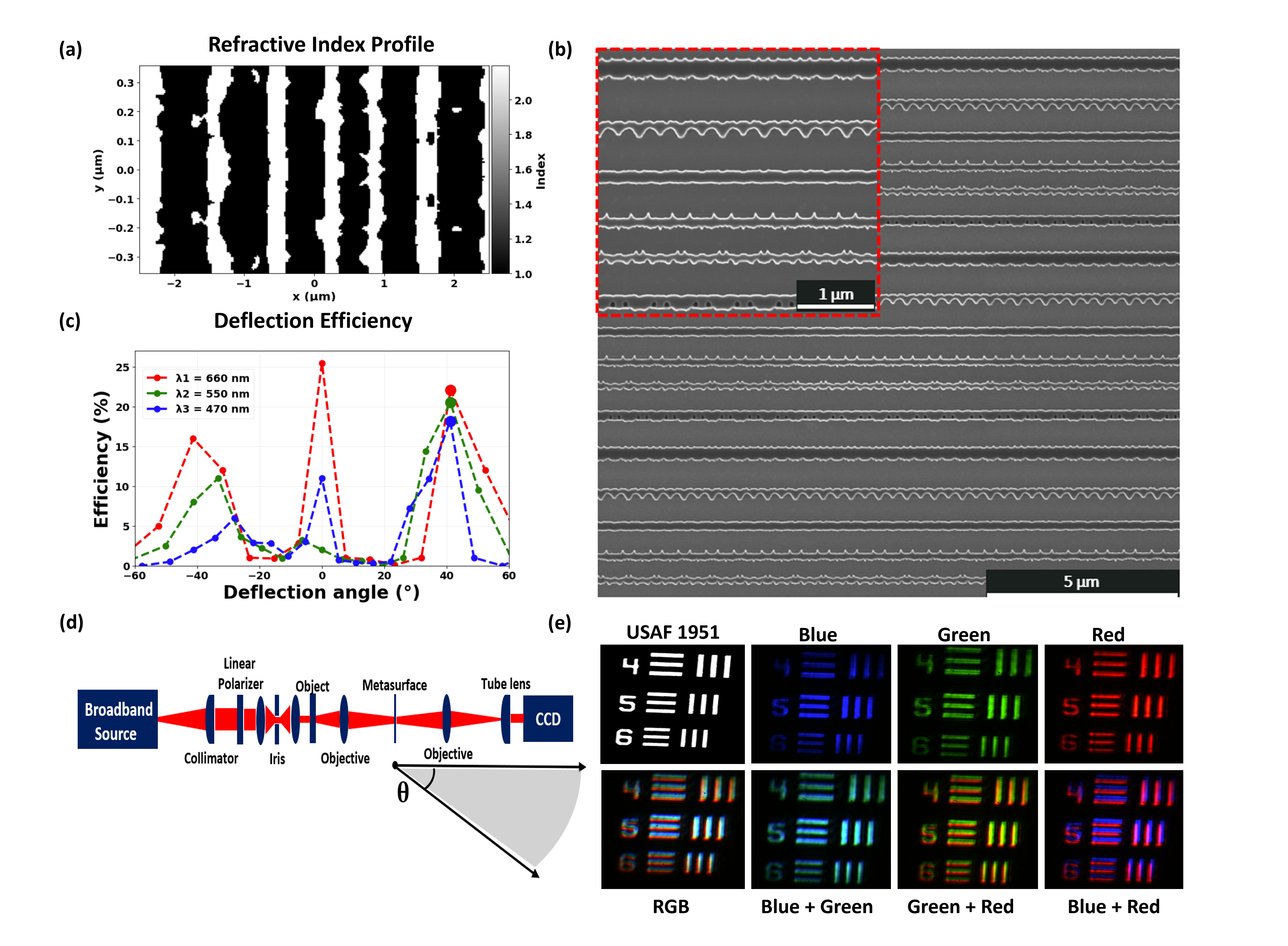}
  \caption{
    \textbf{(a)} Refractive-index profile of the optimized metasurface unit cell.
    \textbf{(b)} SEM of the fabricated metasurface deflector. The inset (red box) highlights the subwavelength corrugations critical for broadband operation. Scale bars: \(5\,\mu\mathrm{m}\) (main image) and \(1\,\mu\mathrm{m}\) (inset).
    \textbf{(c)} Experimentally measured deflection efficiency as a function of output angle for three design wavelengths: 470\,nm (blue), 550\,nm (green), and 660\,nm (red). The prominent efficiency peaks near \(\theta = 41.3^\circ\) confirm the metasurface’s broadband and angle-specific beam deflection performance. The measured efficiencies agree well with theoretical simulations (25.0\% for red , 24.2\% for green, and 24.1\% for blue).
    \textbf{(d)} Schematic of the broadband optical characterization setup. A collimated beam from a broadband source passes through a linear polarizer and imaging optics, illuminates the object and metasurface, and is then focused onto a CCD detector. The system enables angle-resolved measurements, with the detection angle $\theta = 41.3^\circ$ indicated.
    \textbf{(e)} Optical characterization using a USAF 1951 resolution target under illumination at the same three wavelengths and their combinations. Individual and combined RGB (red, green, blue) channel images validate achromatic beam deflection across the visible spectrum.
  }
  \label{fig:fab}
\end{figure}

The characterization system is schematically shown in Figure~\ref{fig:fab}(d). A SuperK FIANIUM broadband source (NKT Photonics) precisely generates red, green, and blue wavelengths at 660~nm, 550~nm, and 470~nm, respectively, forming a multi-wavelength beam. This beam is first linearly polarized by passing through a linear polarizer. To measure the deflection efficiency at each wavelength, the power of the incident plane-wave at that wavelength is measured by a detector placed right before the metasurface. The output power is then measured as a function of the deflected angle and divided by the total incident power to obtain the deflection efficiency at the given wavelength. Figure~\ref{fig:fab}(c) shows the measured deflection efficiency as a function of the angle for the three wavelengths. It is clearly seen that all beams deflect at $41.3^\circ$ with similar efficiencies. The experimentally measured efficiencies—22.1\% for red, 20.5\% for green, and 17.7\% for blue—demonstrate good agreement with theoretical results and confirm robust broadband deflection performance. This represents a major improvement over previously published results~\cite{choi2024multiwavelength}, both in terms of the uniformity of efficiency over a wide spectral range and the absolute diffraction efficiencies. These results highlight the clear advantage of HiLAB over conventional TO in finding an optimal design. The slightly lower experimental efficiency observed for blue light is primarily attributed to material absorption. While the loss of TiO$_2$ at visible wavelengths is generally low, its extinction coefficient $k$ becomes slightly elevated at shorter wavelengths, resulting in increased absorption for blue light. This wavelength-dependent behavior is confirmed by the measured complex refractive index of a TiO$_2$ film, shown in Supplementary Figure~S4, where both $n$ and $k$ are plotted as functions of wavelength. In another experiment, the polarized light illuminates a standard USAF 1951 resolution test target and subsequently passes through an objective lens. The beam then illuminates the fabricated achromatic beam deflector (metasurface), which redirects it toward an imaging arm mounted on a large-area rotating breadboard, enabling precise angular scanning of the output beam. Finally, a tube lens relays the image onto a CCD camera, capturing the test target pattern with spectral and polarization modifications introduced by the metasurface. The output pattern of the metasurface at the desired wavelength is imaged under illumination with the patterned light at different input light wavelengths and their combinations. The results, shown in Figure~\ref{fig:fab}(e), clearly demonstrate the ability of the fabricated device in deflecting different colors at the same angle with almost the same efficiency. 
\subsection{Discussions}
The key aspects of HiLAB are the use of conventional gradient-based optimization techniques in a low-dimensional space and using BO to widen the local search region and avoid early convergence to local optima. The resulting reduction in computation requirements and the ability to search for a close-to-global optimum (or a strong local optimum) make it useful for the design of large-scale metasurfaces and the addition of multifunctional features. Indeed, this focus on complex functionality~\cite{nikkhah2024inverse} aligns with a broader trend in the field, as recent works at the intersection of deep learning and TO using hybrid inverse-design frameworks further highlight, demonstrating intelligent~\cite{qian2025guidance} and adaptable nanophotonic systems~\cite{qian2020deep}. These studies underscore the growing need for hybrid frameworks that combine the expressivity of AI with the rigor of physics-based optimization. Further optimization and refinement of these individual parts can result in dramatic improvement of the performance of HiLAB and enable its application for a wide range of nanophotonic designs that are not currently possible with existing approaches. Nevertheless, despite being at the early stage of development, the theoretical and experimental results presented in this paper clearly demonstrate the unique features of HiLAB and pinpoint the importance of and the need for using hybrid inverse-design techniques that scale up our design capabilities to larger structures while avoiding weak local optima. Further improvements of such approaches are needed to impose more constraints on the optimization process. Example constraints of interest in nanophotonics include: 1) low sensitivity to fabrication imperfections, 2) increasing the minimum required feature sizes, 3) low sensitivity to environmental changes and alignment errors, and so on. Our proposed hybrid inverse-design framework, HiLAB, is highly general and readily adaptable across a wide variety of photonic inverse design problems. Specifically, HiLAB can utilize established gradient-based optimization methods and commercial software tools~\cite{hugonin2021reticolo, jiang2019global, colburn2021inverse, hazineh2022dflat, jiang2020metanet, su2020nanophotonic, tidy3d_flexcompute_2024} to efficiently generate targeted training datasets. These datasets enable effective exploration of the design space through subsequent data-driven approaches.

The transferability of the trained VAE~\cite{bengio2013representation} is a powerful feature of the HiLAB framework, aligning with the development of general design platforms that can adapt to new objectives without complete retraining~\cite{yu2023general}. In our approach, the VAE learns a physics-agnostic representation of the device geometry~\cite{marzban2025inverse}, leaving the task of solving the physics to the numerical solver. This learned representation can therefore be reused without retraining for other applications involving the same device architecture. The usefulness of this transferred representation will depend on how well the learned geometrical insights apply to the new problem. We have demonstrated this by transferring the latent representation from a VAE trained for a 41.3\(^\circ\) beam deflection to a new target of 30\(^\circ\) deflection angle without retraining. The transferred representation yielded a comparable FoM of 0.20, confirming that the learned geometrical features are indeed transferable to closely related tasks. It remains an open question whether this representation will remain equally effective when the design objective changes more drastically (e.g., from a beam deflector to a broadband absorber). In such cases, even if direct transfer is less effective, the pretrained representation may still be valuable as a warm start~\cite{tripathy2018deep}, as it encodes transferable geometric inductive biases that can accelerate optimization or reduce data requirements\cite{xie2019graph}.

\section*{Conclusion}
In this paper, we introduced \emph{HiLAB}, a novel hybrid inverse-design framework that effectively combines partial TO, generative AI via a ViT-based VAE, and BO to efficiently address complex multifunctional photonic device design challenges. By truncating traditional adjoint-driven TO at early stages and systematically varying key physical parameters, we generate diverse and computationally affordable datasets that capture the critical structural features necessary for high-performance device design. The trained VAE encodes these partially optimized structures into a compact latent space, significantly reducing computational demands for subsequent design exploration. By leveraging BO within this learned latent space, \emph{HiLAB} systematically guides the search towards global optima (or strong local optima) while simultaneously optimizing physical hyperparameters such as device thickness and period. We demonstrated the effectiveness of \emph{HiLAB} by designing an achromatic beam deflector achieving balanced diffraction efficiencies, theoretically exceeding 24\% across red, green, and blue wavelengths—substantially outperforming similar devices designed by prior conventional TO methods and significantly reducing chromatic aberrations. Crucially, the generality and flexibility of \emph{HiLAB} mean it can be easily adapted for different design targets, constraints, or fabrication tolerances without necessitating extensive additional simulations or retraining. By providing a systematic, scalable, and efficient inverse-design methodology, \emph{HiLAB} holds considerable promise for accelerating the development of robust, high-performance photonic and metamaterial devices across diverse application domains.

\medskip
\section*{Supporting Information}
Correspondence and requests for materials should be addressed to A. Adibi.
\medskip
\section*{Acknowledgements}
The authors gratefully acknowledge support from the Georgia Institute of Technology's Institute for Matter and Systems (IMS). This work was supported by the National Science Foundation.  
\bibliographystyle{unsrt}
\bibliography{refrences}

\begin{thebibliography}{10}

\bibitem{kuznetsov2024roadmap}
Arseniy~I Kuznetsov, Mark~L Brongersma, Jin Yao, Mu~Ku Chen, Uriel Levy, Din~Ping Tsai, Nikolay~I Zheludev, Andrei Faraon, Amir Arbabi, Nanfang Yu, et~al.
\newblock Roadmap for optical metasurfaces.
\newblock {\em ACS photonics}, 11(3):816--865, 2024.

\bibitem{li2022empowering}
Zhaoyi Li, Rapha{\"e}l Pestourie, Zin Lin, Steven~G Johnson, and Federico Capasso.
\newblock Empowering metasurfaces with inverse design: principles and applications.
\newblock {\em Acs Photonics}, 9(7):2178--2192, 2022.

\bibitem{lin2022end}
Zin Lin, Rapha{\"e}l Pestourie, Charles Roques-Carmes, Zhaoyi Li, Federico Capasso, Marin Solja{\v{c}}i{\'c}, and Steven~G Johnson.
\newblock End-to-end metasurface inverse design for single-shot multi-channel imaging.
\newblock {\em Optics express}, 30(16):28358--28370, 2022.

\bibitem{pestourie2020assume}
Rapha{\"e}l Jean-Marie~Fernand Pestourie.
\newblock {\em Assume your neighbor is your equal: Inverse design in nanophotonics}.
\newblock Harvard University, 2020.

\bibitem{molesky2018inverse}
Sean Molesky, Zin Lin, Alexander~Y Piggott, Weiliang Jin, Jelena Vuckovi{\'c}, and Alejandro~W Rodriguez.
\newblock Inverse design in nanophotonics.
\newblock {\em Nature Photonics}, 12(11):659--670, 2018.

\bibitem{sell2017large}
David Sell, Jianji Yang, Sage Doshay, Rui Yang, and Jonathan~A Fan.
\newblock Large-angle, multifunctional metagratings based on freeform multimode geometries.
\newblock {\em Nano letters}, 17(6):3752--3757, 2017.

\bibitem{jiang2019global}
Jiaqi Jiang and Jonathan~A Fan.
\newblock Global optimization of dielectric metasurfaces using a physics-driven neural network.
\newblock {\em Nano letters}, 19(8):5366--5372, 2019.

\bibitem{gershnabel2022reparameterization}
Erez Gershnabel, Mingkun Chen, Chenkai Mao, Evan~W Wang, Philippe Lalanne, and Jonathan~A Fan.
\newblock Reparameterization approach to gradient-based inverse design of three-dimensional nanophotonic devices.
\newblock {\em ACS Photonics}, 10(4):815--823, 2022.

\bibitem{li2022inverse}
Zhaoyi Li, Rapha{\"e}l Pestourie, Joon-Suh Park, Yao-Wei Huang, Steven~G Johnson, and Federico Capasso.
\newblock Inverse design enables large-scale high-performance meta-optics reshaping virtual reality.
\newblock {\em Nature communications}, 13(1):2409, 2022.

\bibitem{lin2021end}
Zin Lin, Charles Roques-Carmes, Rapha{\"e}l Pestourie, Marin Solja{\v{c}}i{\'c}, Arka Majumdar, and Steven~G Johnson.
\newblock End-to-end nanophotonic inverse design for imaging and polarimetry.
\newblock {\em Nanophotonics}, 10(3):1177--1187, 2021.

\bibitem{muhammad2025second}
Naseer Muhammad, Azra Begum, Zhaoxian Su, and Lingling Huang.
\newblock Second harmonic generation from bound-state in the continuum-hosted few-layers van der waals metasurface.
\newblock {\em Nanophotonics}, 14(2):263--270, 2025.

\bibitem{fu2024optical}
Tingzhao Fu, Jianfa Zhang, Run Sun, Yuyao Huang, Wei Xu, Sigang Yang, Zhihong Zhu, and Hongwei Chen.
\newblock Optical neural networks: progress and challenges.
\newblock {\em Light: Science \& Applications}, 13(1):263, 2024.

\bibitem{zarei2020integrated}
Sanaz Zarei, Mahmood-reza Marzban, and Amin Khavasi.
\newblock Integrated photonic neural network based on silicon metalines.
\newblock {\em Optics Express}, 28(24):36668--36684, 2020.

\bibitem{poordashtban2023integrated}
Omid Poordashtban, Mahmood~Reza Marzabn, and Amin Khavasi.
\newblock Integrated photonic convolutional neural network based on silicon metalines.
\newblock {\em IEEE Access}, 11:61728--61737, 2023.

\bibitem{lin2019topology}
Zin Lin, Victor Liu, Rapha{\"e}l Pestourie, and Steven~G Johnson.
\newblock Topology optimization of freeform large-area metasurfaces.
\newblock {\em Optics express}, 27(11):15765--15775, 2019.

\bibitem{bendsoe2013topology}
Martin~Philip Bendsoe and Ole Sigmund.
\newblock {\em Topology optimization: theory, methods, and applications}.
\newblock Springer Science \& Business Media, 2013.

\bibitem{yu2025angle}
Haoxiang Yu, Xuxi Zhou, Yunlai Fu, Yuanqing Wan, Weijun Liu, Quan Yuan, and Shuming Wang.
\newblock Angle-insensitive broadband multispectro-polarimetric encoding based on inverse design of mosaic metasurfaces.
\newblock {\em Small Methods}, page 2402182, 2025.

\bibitem{liao2024low}
Junpeng Liao, Dongmei Huang, Yegang Lu, Yan Li, and Ye~Tian.
\newblock Low-loss and compact arbitrary-order silicon mode converter based on hybrid shape optimization.
\newblock {\em Nanophotonics}, 13(22):4137--4148, 2024.

\bibitem{dainese2024shape}
Paulo Dainese, Louis Marra, Davide Cassara, Ary Portes, Jaewon Oh, Jun Yang, Alfonso Palmieri, Janderson~Rocha Rodrigues, Ahmed~H Dorrah, and Federico Capasso.
\newblock Shape optimization for high efficiency metasurfaces: theory and implementation.
\newblock {\em Light: Science \& Applications}, 13(1):300, 2024.

\bibitem{kiarashinejad2020deep}
Yashar Kiarashinejad, Sajjad Abdollahramezani, and Ali Adibi.
\newblock Deep learning approach based on dimensionality reduction for designing electromagnetic nanostructures.
\newblock {\em npj Computational Materials}, 6(1):12, 2020.

\bibitem{liu2018generative}
Zhaocheng Liu, Dayu Zhu, Sean~P Rodrigues, Kyu-Tae Lee, and Wenshan Cai.
\newblock Generative model for the inverse design of metasurfaces.
\newblock {\em Nano letters}, 18(10):6570--6576, 2018.

\bibitem{an2019deep}
Sensong An, Clayton Fowler, Bowen Zheng, Mikhail~Y Shalaginov, Hong Tang, Hang Li, Li~Zhou, Jun Ding, Anuradha~Murthy Agarwal, Clara Rivero-Baleine, et~al.
\newblock A deep learning approach for objective-driven all-dielectric metasurface design.
\newblock {\em Acs Photonics}, 6(12):3196--3207, 2019.

\bibitem{hemmatyar2019full}
Omid Hemmatyar, Sajjad Abdollahramezani, Yashar Kiarashinejad, Mohammadreza Zandehshahvar, and Ali Adibi.
\newblock Full color generation with fano-type resonant hfo 2 nanopillars designed by a deep-learning approach.
\newblock {\em Nanoscale}, 11(44):21266--21274, 2019.

\bibitem{zandehshahvar2023metric}
Mohammadreza Zandehshahvar, Yashar Kiarashi, Muliang Zhu, Daqian Bao, Mohammad H~Javani, Reza Pourabolghasem, and Ali Adibi.
\newblock Metric learning: harnessing the power of machine learning in nanophotonics.
\newblock {\em Acs Photonics}, 10(4):900--909, 2023.

\bibitem{pestourie2023physics}
Rapha{\"e}l Pestourie, Youssef Mroueh, Chris Rackauckas, Payel Das, and Steven~G Johnson.
\newblock Physics-enhanced deep surrogates for partial differential equations.
\newblock {\em Nature Machine Intelligence}, 5(12):1458--1465, 2023.

\bibitem{li2022ultracompact}
Yan Li, Shuyi Chen, Haowen Liang, Xiuying Ren, Lingcong Luo, Yuye Ling, Shuxin Liu, Yikai Su, and Shin-Tson Wu.
\newblock Ultracompact multifunctional metalens visor for augmented reality displays.
\newblock {\em PhotoniX}, 3(1):29, 2022.

\bibitem{bonod2016diffraction}
Nicolas Bonod and J{\'e}r{\^o}me Neauport.
\newblock Diffraction gratings: from principles to applications in high-intensity lasers.
\newblock {\em Advances in Optics and Photonics}, 8(1):156--199, 2016.

\bibitem{choi2024multiwavelength}
Taewon Choi, Chulsoo Choi, Junseo Bang, Youngjin Kim, Hyunwoo Son, Changhyun Kim, Junhyeok Jang, Yoonchan Jeong, and Byoungho Lee.
\newblock Multiwavelength achromatic deflector in the visible using a single-layer freeform metasurface.
\newblock {\em Nano Letters}, 24(35):10980--10986, 2024.

\bibitem{sell2019adjoint}
David Sell.
\newblock {\em Adjoint Optimization of Free-space Metasurfaces}.
\newblock Stanford University, 2019.

\bibitem{pestourie2025fastapproximatesolversmetamaterials}
Raphael Pestourie.
\newblock Fast approximate solvers for metamaterials design in electromagnetism, 2025.

\bibitem{mordvintsev2014opencv}
Alexander Mordvintsev and K~Abid.
\newblock Opencv-python tutorials documentation.
\newblock {\em Obtenido de https://media. readthedocs. org/pdf/opencv-python-tutroals/latest/opencv-python-tutroals. pdf}, 2014.

\bibitem{su2020nanophotonic}
Logan Su, Dries Vercruysse, Jinhie Skarda, Neil~V Sapra, Jan~A Petykiewicz, and Jelena Vu{\v{c}}kovi{\'c}.
\newblock Nanophotonic inverse design with spins: Software architecture and practical considerations.
\newblock {\em Applied Physics Reviews}, 7(1), 2020.

\bibitem{christiansen2021inverse}
Rasmus~E Christiansen and Ole Sigmund.
\newblock Inverse design in photonics by topology optimization: tutorial.
\newblock {\em Journal of the Optical Society of America B}, 38(2):496--509, 2021.

\bibitem{doersch2016tutorial}
Carl Doersch.
\newblock Tutorial on variational autoencoders.
\newblock {\em arXiv preprint arXiv:1606.05908}, 2016.

\bibitem{dosovitskiy2020image}
Alexey Dosovitskiy, Lucas Beyer, Alexander Kolesnikov, Dirk Weissenborn, Xiaohua Zhai, Thomas Unterthiner, Mostafa Dehghani, Matthias Minderer, Georg Heigold, Sylvain Gelly, et~al.
\newblock An image is worth 16x16 words: Transformers for image recognition at scale.
\newblock {\em arXiv preprint arXiv:2010.11929}, 2020.

\bibitem{wu2020visual}
Bichen Wu, Chenfeng Xu, Xiaoliang Dai, Alvin Wan, Peizhao Zhang, Zhicheng Yan, Masayoshi Tomizuka, Joseph Gonzalez, Kurt Keutzer, and Peter Vajda.
\newblock Visual transformers: Token-based image representation and processing for computer vision, 2020.

\bibitem{deng2009imagenet}
Jia Deng, Wei Dong, Richard Socher, Li-Jia Li, Kai Li, and Li~Fei-Fei.
\newblock Imagenet: A large-scale hierarchical image database.
\newblock In {\em 2009 IEEE conference on computer vision and pattern recognition}, pages 248--255. Ieee, 2009.

\bibitem{howard2018universal}
Jeremy Howard and Sebastian Ruder.
\newblock Universal language model fine-tuning for text classification.
\newblock {\em arXiv preprint arXiv:1801.06146}, 2018.

\bibitem{zeiler2014visualizing}
Matthew~D Zeiler and Rob Fergus.
\newblock Visualizing and understanding convolutional networks.
\newblock In {\em European conference on computer vision}, pages 818--833. Springer, 2014.

\bibitem{ridnik2021imagenet}
Tal Ridnik, Emanuel Ben-Baruch, Asaf Noy, and Lihi Zelnik-Manor.
\newblock Imagenet-21k pretraining for the masses.
\newblock {\em arXiv preprint arXiv:2104.10972}, 2021.

\bibitem{kingma2013auto}
Diederik~P Kingma, Max Welling, et~al.
\newblock Auto-encoding variational bayes, 2013.

\bibitem{ioffe2015batch}
Sergey Ioffe and Christian Szegedy.
\newblock Batch normalization: Accelerating deep network training by reducing internal covariate shift.
\newblock In {\em International conference on machine learning}, pages 448--456. pmlr, 2015.

\bibitem{agarap2018deep}
Abien~Fred Agarap.
\newblock Deep learning using rectified linear units (relu).
\newblock {\em arXiv preprint arXiv:1803.08375}, 2018.

\bibitem{kirkland2010bilinear}
Earl~J Kirkland.
\newblock Bilinear interpolation.
\newblock In {\em Advanced computing in electron microscopy}, pages 261--263. Springer, 2010.

\bibitem{gramacy2020surrogates}
Robert~B Gramacy.
\newblock {\em Surrogates: Gaussian process modeling, design, and optimization for the applied sciences}.
\newblock Chapman and Hall/CRC, 2020.

\bibitem{mcintire2016sparse}
Mitchell McIntire, Daniel Ratner, and Stefano Ermon.
\newblock Sparse gaussian processes for bayesian optimization.
\newblock In {\em UAI}, volume~16, pages 517--526, 2016.

\bibitem{xi2023ultrahigh}
Wang Xi, Yun-Jo Lee, Shilv Yu, Zihe Chen, Junichiro Shiomi, Sun-Kyung Kim, and Run Hu.
\newblock Ultrahigh-efficient material informatics inverse design of thermal metamaterials for visible-infrared-compatible camouflage.
\newblock {\em Nature Communications}, 14(1):4694, 2023.

\bibitem{chen2023multi}
Zihe Chen, Shilv Yu, Bin Hu, and Run Hu.
\newblock Multi-band and wide-angle nonreciprocal thermal radiation.
\newblock {\em International Journal of Heat and Mass Transfer}, 209:124149, 2023.

\bibitem{seeger2004gaussian}
Matthias Seeger.
\newblock Gaussian processes for machine learning.
\newblock {\em International journal of neural systems}, 14(02):69--106, 2004.

\bibitem{nikkhah2024inverse}
Vahid Nikkhah, Ali Pirmoradi, Farshid Ashtiani, Brian Edwards, Firooz Aflatouni, and Nader Engheta.
\newblock Inverse-designed low-index-contrast structures on a silicon photonics platform for vector--matrix multiplication.
\newblock {\em Nature Photonics}, 18(5):501--508, 2024.

\bibitem{qian2025guidance}
Chao Qian, Ido Kaminer, and Hongsheng Chen.
\newblock A guidance to intelligent metamaterials and metamaterials intelligence.
\newblock {\em Nature Communications}, 16(1):1154, 2025.

\bibitem{qian2020deep}
Chao Qian, Bin Zheng, Yichen Shen, Li~Jing, Erping Li, Lian Shen, and Hongsheng Chen.
\newblock Deep-learning-enabled self-adaptive microwave cloak without human intervention.
\newblock {\em Nature photonics}, 14(6):383--390, 2020.

\bibitem{hugonin2021reticolo}
Jean~Paul Hugonin and Philippe Lalanne.
\newblock Reticolo software for grating analysis.
\newblock {\em arXiv preprint arXiv:2101.00901}, 2021.

\bibitem{colburn2021inverse}
Shane Colburn and Arka Majumdar.
\newblock Inverse design and flexible parameterization of meta-optics using algorithmic differentiation.
\newblock {\em Communications Physics}, 4(1):65, 2021.

\bibitem{hazineh2022dflat}
Dean~S. Hazineh, Soon Wei~Daniel Lim, Zhujun Shi, Federico Capasso, Todd Zickler, and Qi~Guo.
\newblock D-flat: A differentiable flat-optics framework for end-to-end metasurface visual sensor design, 2022.

\bibitem{jiang2020metanet}
Jiaqi Jiang, Robert Lupoiu, Evan~W Wang, David Sell, Jean Paul~Hugonin, Philippe Lalanne, and Jonathan~A Fan.
\newblock Metanet: a new paradigm for data sharing in photonics research.
\newblock {\em Optics express}, 28(9):13670--13681, 2020.

\bibitem{tidy3d_flexcompute_2024}
{Tidy3D, the full-wave simulator employing the finite-difference time-domain (FDTD) method, developed by Flexcompute, Inc}.
\newblock https://www.flexcompute.com/tidy3d/solver, (2024).

\bibitem{bengio2013representation}
Yoshua Bengio, Aaron Courville, and Pascal Vincent.
\newblock Representation learning: A review and new perspectives.
\newblock {\em IEEE transactions on pattern analysis and machine intelligence}, 35(8):1798--1828, 2013.

\bibitem{yu2023general}
Shilv Yu, Peng Zhou, Wang Xi, Zihe Chen, Yuheng Deng, Xiaobing Luo, Wangnan Li, Junichiro Shiomi, and Run Hu.
\newblock General deep learning framework for emissivity engineering.
\newblock {\em Light: Science \& Applications}, 12(1):291, 2023.

\bibitem{marzban2025inverse}
Reza Marzban, Ali Adibi, and Raphael Pestourie.
\newblock Inverse design in nanophotonics via representation learning.
\newblock {\em arXiv preprint arXiv:2507.00546}, 2025.

\bibitem{tripathy2018deep}
Rohit~K Tripathy and Ilias Bilionis.
\newblock Deep uq: Learning deep neural network surrogate models for high dimensional uncertainty quantification.
\newblock {\em Journal of computational physics}, 375:565--588, 2018.

\bibitem{xie2019graph}
Tian Xie, Arthur France-Lanord, Yanming Wang, Yang Shao-Horn, and Jeffrey~C Grossman.
\newblock Graph dynamical networks for unsupervised learning of atomic scale dynamics in materials.
\newblock {\em Nature communications}, 10(1):2667, 2019.

\end{thebibliography}
\end{document}